\documentstyle[proceedings,psfig]{crckapb}

\def\msun{$M_\odot$}
\def\about{$\sim$}
\def\mdot{$\dot M$}
\def\approxlt{\ifmmode \rlap{$<$}{}_{{}_{{}_{\textstyle\sim}}} \else%
$\ifmmode \rlap{$<$}{}_{{}_{{}_{\textstyle\sim}}}$\fi}
\def\da{$\downarrow$}
\def\mc{\multicolumn}

\begin{opening}
\title{KILOHERTZ QUASI-PERIODIC OSCILLATIONS IN LOW-MASS X-RAY BINARIES}
\author{M. van der Klis}
\institute{Astronomical Institute ``Anton Pannekoek'' \\ University of
Amsterdam \\ Kruislaan 403, 1098 SL Amsterdam, The Netherlands}
\end{opening}
\runningtitle{Kilohertz QPO in LMXBs}
\begin{document}
\begin{abstract}
In early 1996 a series of discoveries begun with NASA's Rossi X-ray
Timing Explorer of a new, up to then unknown astrophysical
phenomenon. It turned out that accreting low magnetic-field neutron
stars show quasi-periodic oscillations in their X-ray flux at rates of
up to more than a kilohertz. These kHz QPO, now reported from eleven
different systems, are among the fastest phenomena in the sky and can
provide us with new information about the fundamental properties of
neutron stars and help testing general relativity in the strong-field
regime. If, for example, their frequencies can be identified with the
Keplerian frequencies of matter in orbit around a 1.4\msun\ neutron
star, then the radius of the star would have to be less than 15\,km,
which directly constrains the equation of state of bulk
nuclear-density matter, and for an only slightly tighter orbit or
slightly more massive neutron star the orbital radius would equal the
Schwarzschild-geometry general-relativistic marginally stable orbit
(12.5\,km for a 1.4\msun\ object). So far all models that have been
put forward for explaining the new phenomenon have encountered
problems. In this paper I review the relatively simple and highly
suggestive phenomenology as it has emerged from the data up to now,
and discuss some of the proposed models.
\end{abstract}

\section{Introduction}

The main
motivation for studying X-ray binaries is not that they exhibit a wide
range of complex phenomenology, which they do, but that they contain
neutron stars (and black holes), objects of fundamental physical
interest, and allow to derive information about the equation of state
of high-density matter and perform tests of general relativity in the
strong-field regime. In this talk, I shall be discussing low-mass
X-ray binaries (LMXBs) containing neutron stars exclusively, as it is
in the understanding of the physics of these systems that great
progress has recently become possible by the discovery, with NASA's
Rossi X-ray Timing Explorer (RXTE), of a new phenomenon, kilohertz
quasi-periodic oscillations (kHz QPO).

In these X-ray binary systems matter is transferred from a low-mass
(\approxlt1\msun) star to a neutron star by way of an accretion
disk. The X-rays originate from the hot (\about10$^7$\,K) plasma
comprising the inner few 10$^1$ kilometers of the flow. This is very
close to the neutron star, which itself has a radius, $R$, of order
10\,km, so that by studying the properties of this flow one expects to
be able to derive information about the star.

The high temperatures in the inner flow are caused by the release of
large amounts of graviational energy when the matter descends into the
neutron star's very deep gravitational potential well
($GM/R\sim0.2c^2$; here and below I assume $M=1.4$\msun\ for the
neutron star's mass). The characteristic velocities near the star are
of order $(GM/R)^{1/2}\sim0.5c$. Therefore the dynamical time scale,
the time scale for motion of matter through the emitting region, is
short; $\tau_{dyn} \equiv (r^3/GM)^{1/2}$$\sim$0.1\,ms for $r$=10\,km,
and \about2\,ms for $r$=100\,km.

Up to less than a year ago, no direct information existed about the
properties of these flows at these time scales. In this paper I report
on how, since February 1996, we are for the first time actually
observing time variability from accretion flows onto neutron stars at
the expected millisecond time scales. A new rapid-variability
phenomenon has been discovered, namely quasi-periodic oscillations in
the X-ray flux with amplitudes of up to several 10\% of the total
flux, quality factors $Q\equiv\Delta\nu/\nu$ (see \S2) of up to
several 100, and frequencies of up to \about1200\,Hz. I shall call
this phenomenon ``kHz QPO'' (kilohertz quasi-periodic oscillations)
throughout the rest of this paper.

A great deal of information is available about the properties of LMXBs
and the physics of accretion onto a neutron star. The last pre-kHz-QPO
overview of rapid X-ray variability in X-ray binaries can be found in
the Lewin et al. book ``X-Ray Binaries'' (van der Klis 1995; look here
if you wish to find out about atoll sources, Z sources and the
latters' 16--60\,Hz horizontal-branch oscillations and the 6--20\,Hz
normal-flaring branch oscillations). For understanding what follows,
it is useful to remind the reader of the usual terminology with
respect to the subclasses of LMXBs (Hasinger and van der Klis 1989): Z
sources are near-Eddington accretors and probably have somewhat
stronger (1--5 10$^9$\,G) magnetic fields, atoll sources are often X-ray
burst sources, have luminosities between 10$^{-3}$\,L$_{Edd}$ and a
few 10$^{-1}$\,L$_{Edd}$, and are thought to have somewhat weaker
magnetic fields (10$^8$--10$^9$\,G).

X-ray astronomers are presently scrambling to try and make sense of
the phenomenology of kHz QPO, which turn out to be at the same time
highly suggestive of interpretation and very restrictive of possible
models, and theorists have already begun working out sophisticated
models. None of this has reached an equilibrium state yet, and what I
report in this paper will necessarily be of a ``snapshot''
nature. What is clear at this point is that for the first time we are
seeing a rapid X-ray variability phenomenon that is directly linked
with a neutron star's most distinguishing characteristic (only shared
among macroscopic objects with stellar-mass black holes): its
compactness. This is particularly evident if the phenomena are in some
way related to orbital motion.  After all, a Keplerian orbital
frequency $\nu_K = P_{orb}^{-1}=(GM/4\pi^2r_K^3)^{1/2}$ of 1200\,Hz
around a 1.4\msun\ neutron star as seen from infinity corresponds to
an orbital radius $r_K = (GM/4\pi^2\nu_K^2)^{1/3}$ of 15\,km, directly
constraining the equation of state of the bulk nuclear-density matter,
and only just outside the general-relativistic marginally stable
orbit. Whatever the model, for the first time we have to seriously
worry about general-relativistic effects in describing the observable
dynamics of the physical system.

\section{Observations and interpretation}

Kilohertz QPO have now\footnote{March 24, 1997} been reported from 11
LMXBs, 3 of which are Z sources and 8 of which are atoll sources and
probable atoll sources (see van der Klis 1995; hereafter I shall use
``atoll source'' for LMXBs that probably fall in this class as well as
for those that definitely do so), together covering nearly three
orders of magnitude in X-ray luminosity (\about10$^{-3}$ to
\about1\,L$_{Edd}$). Table\,1 summarizes some of these results, and
provides an overview of the literature that is approximately complete
as of this writing. Rather than getting into an exhaustive description
of the phenomenology or following the historical line, I shall
concentrate on what I consider at this point to be the main clues. I
refer to the Table for all kHz QPO observational references in the
remainder of this section.

\begin{table}[htb]
\begin{center}\scriptsize\parskip=0pt\baselineskip=8pt
\renewcommand{\arraystretch}{0.7}\setlength{\doublerulesep}{1pt}
\caption{Observed frequencies of kilohertz QPO.} \label{table1}
\begin{tabular}{lccccl}
Source          & Lower  & Upper   & Peak       & ``Third''   & References\\
(in order       & peak   & peak    & sepa-      & freq.\\
of RA)          & freq.  & freq.   & ration     &     \\
                & (Hz)   & (Hz)    & (Hz)       & (Hz)\\ 
\hline
\noalign{\vskip0.2mm}
4U\,0614+091    &        & 520     &            &             & Ford et al. 1996, 1997 \\
                &        & \da     &            &             & van der Klis et al. 1996d \\
                & 480    & 750     &            &             & Mendez et al. 1997 \\
                & \da    & \da     & 327$\pm$4  & 328         & Vaughan et al. 1997 \\
                & 800    & 1150    &            &             & \\
\hline
\noalign{\vskip0.2mm}
4U\,1608$-$52   & \mc{2}{c}{\ 691} &            &             & Van Paradijs et al. 1996\\
                & \mc{2}{c}{\ 830} &            &             & Berger et al. 1996\\
                & \mc{2}{c}{\ \da} &            &             & Vaughan et al. 1997 \\
                & \mc{2}{c}{\ 890} &            &             & \\
\hline
\noalign{\vskip0.2mm}
Sco\,X-1        & 570    & 870     &            &             & van der Klis et al. 1996a,b,c, \\
                & \da    & \da     & 292$\pm$2  &             & 1997b\\
                & 800    & 1050    &\da         &             & \\
                & \da    & \da     & 247$\pm$3  &             & \\
                & 830    & 1080    &            &             & \\
                &        & \da     &            &             & \\
                &        & 1130    &            &             & \\
\hline
\noalign{\vskip0.2mm}
4U\,1636$-$53   & 898    & 1147    &            &             & Zhang et al.  1996, 1997 \\ 
                & \da    & \da     & 249$\pm$13 & 581         & van der Klis et al. 1996d \\ 
                & 920    & 1183    &            &             & Wijnands et al. 1997 \\ 
                &        & \da     &            &             & Vaughan et al. 1997 \\ 
                &        & 1193    &            &             & \\      
\noalign{\vskip0.2mm}
                & \mc{2}{c}{\ 835} &            &             & \\
                & \mc{2}{c}{\ \da }&            &             & \\
                & \mc{2}{c}{\ 897} &            &             & \\
\hline
\noalign{\vskip0.2mm}
4U\,1728$-$34   &        &  500    &            &             & Strohmayer et al. 1996a,b,c \\
                &        & \da     &            &             & \\
                & 640    &  990    &            &             & \\
                & \da    & \da     & 355$\pm$5  & 363         & \\
                & 790    & 1100    &            &             & \\
\hline
\noalign{\vskip0.2mm}
KS\,1731$-$260  & 898    & 1159    & 260$\pm$10 & 524         & Morgan
and Smith 1996 \\
                &        & \da     &            &             & Smith et al. 1997 \\
                &        & 1207    &            &             & Wijnands and van der Klis 1997 \\
\hline
\noalign{\vskip0.2mm}
4U\,1735-44     & \mc{2}{c}{\ 1150}&            &             & Wijnands et al. 1996 \\ 
\hline
\noalign{\vskip0.2mm}
X\,1743-29?     &        &         &            & 589         & Strohmayer et al. 1996d \\
\hline
\noalign{\vskip0.2mm}
GX\,5$-$1       &        & 567     &            &             & van der Klis et al. 1996e \\
                &        & \da     &            &             & \\
                & 325    & 652     &            &             & \\
                & \da    & \da     & 327$\pm$11 &             & \\
                & 448    & 746     &            &             & \\
                &        & \da     &            &             & \\
                &        & 895     &            &             & \\
\hline
\noalign{\vskip0.2mm}
GX\,17+2        & 682    & 988     & 306$\pm$5  &             & van der Klis et al. 1997a \\
                & \da    &         &            &             & \\
                & 880    &         &            &             & \\
\hline
\noalign{\vskip0.2mm}
4U\,1820$-$30   & 546    &         &            &             & Smale et al. 1996, 1997 \\
                & \da    &         &            &             & \\
                & 796    & 1065    & 275$\pm$8  &             & \\
\hline
\end{tabular}
Arrows indicate observed frequency variations.\hfill\break
Frequencies in the same row were observed simultaneously, except
``third'' frequencies.\hfill\break
Entries straddling the upper and lower peak columns are of single,
unidentified peaks.\hfill\break
\end{center}
\end{table}

A clear pattern of systematic behaviour has emerged. In most sources
(8 out of 11) two simultaneous kHz peaks (hereafter: twin peaks) are
observed in the power spectra of the X-ray count rate variations
(Fig.\,1). The lower-frequency peak (hereafter the {\it lower peak})
has been observed at frequencies between 325 and 920\,Hz, the
higher-frequency peak (hereafter the {\it upper peak}) has been
observed at frequencies between 500 and 1207\,Hz. When the accretion
rate \mdot\ increases, both peaks move to higher frequency. In atoll
sources \mdot\ is inferred to correlate with X-ray count rate, and kHz
QPO frequency increases with count rate. In Z sources in their
so-called ``normal branch'' (NB),
\mdot\ is inferred to {\it anti}correlate to count rate, and indeed in Z
sources in the NB kHz QPO frequency increases when the count rate
drops.

\begin{figure}[t]
\begin{center}
\begin{tabular}{c}
\psfig{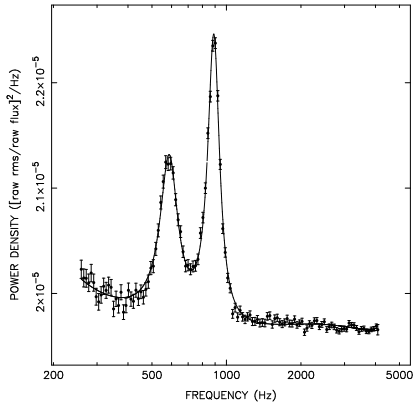}
\end{tabular}
\caption{\scriptsize Power spectrum of Sco\,X-1  
showing double kHz QPO peaks (van der Klis et al. 1997). The sloping
continuum above 1\,kHz is instrumental. \label{fig1}}
\end{center}
\end{figure}

In three atoll sources (4U\,1728$-$34, 4U\,1636$-$53 and
KS\,1731$-$260), ``third peak'' oscillations have been seen during
X-ray bursts whose frequencies (360--580\,Hz) are consistent with
being equal to the frequency {\it differences} between the twin peaks
(in 4U\,1728$-$34), or twice that (in the other two sources). In a
fourth atoll source (4U\,0614+09) there is marginal evidence for a
third peak at the twin-peak separation frequency which corresponds to
an oscillation in the persistent emission rather than in X-ray
bursts. 

These cases of three commensurate frequencies very strongly suggest
that some kind of beat-frequency model is at work, with the ``third
peaks'' at the neutron star spin frequencies (or twice that), the
upper kHz peak at the Kepler frequency corresponding to some preferred
orbital radius around the neutron star, and the lower kHz peak at the
difference frequency between these two.  Strohmayer et al. (1996c)
suggested that this preferred radius is the magnetospheric
radius. Miller, Lamb and Psaltis (1996) proposed it is the sonic
radius. In models of this kind, which involve the neutron-star spin as
one of the frequencies participating in the beat-frequency process,
the twin-peak separation is predicted to be constant. However, in
Sco\,X-1 the peak separation varies systematically with inferred
\mdot, from \about310\,Hz when the upper peak is near 870\,Hz 
to \about230\,Hz when it is near 1075\,Hz: the peaks move closer
together by \about80\,Hz while they both move up in frequency as
\mdot\ increases. This is in strong contradiction to 
straightforward beat-frequency models (see \S3).

In the Z sources Sco\,X-1, GX\,5$-$1 and GX\,17+2 twin kHz QPO peaks
and the so-called horizontal-branch oscillations (HBO; van der Klis et
al. 1985) are seen simultaneously (Fig.\,2). HBO are thought to be
a product of the magnetospheric beat-frequency mechanism (Alpar and
Shaham 1985, Lamb et al. 1985). If this is correct, then this model
can {\it not} explain the kHz QPO in these sources. It is possible in
principle that the kHz QPO in the Z sources is a different phenomenon
from that in the atoll sources (e.g., Strohmayer et al. 1996c), but
this seems unlikely: the frequencies, their dependence on \mdot, the
coherencies, the peak separations and the fact that there are {\it
two} peaks, one of which sometimes becomes undetectable at extreme
\mdot, are too similar to attribute to just coincidence. If this is 
correct, then the variable twin-peak separation seen in Sco\,X-1,
the simultaneous presence of kHz QPO and HBO in Z sources, {\it and}
the direct indications for a beat frequency in the atoll sources must
all be explained within the same model, a formidable challenge.

\begin{figure}[tbp]
\begin{center}
\begin{tabular}{c}
\psfig{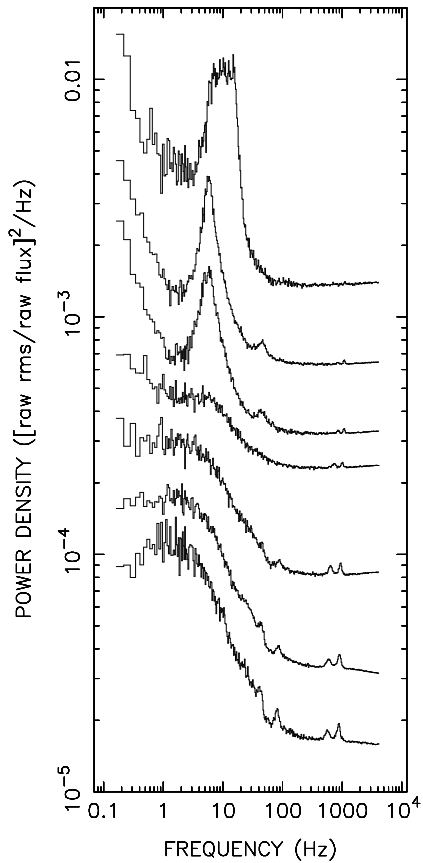}
\end{tabular}
\caption{\scriptsize Power spectra of Sco X-1, with inferred \mdot\ 
increasing upwards. Notice the decrease in strength and increase in
frequency of the kHz peaks as a function of \mdot. The peaks near 45
and 90\,Hz are identified as horizontal branch oscillations (HBO),
that between 6 and 20\,Hz as normal/flaring branch oscillations
(N/FBO). The large width of the N/FBO peak in the top trace is due to
peak motion. The sloping continua in the kHz range are
instrumental. \label{fig2}}
\end{center}
\end{figure}

One of the distinguishing characteristics of kHz QPO is that they
often show a relatively large coherence. The quality factor Q, defined
as the QPO peak's centroid frequency $\nu$ divided by its full width
at half maximum $\Delta\nu$ regularly reaches values of more than 100
in one or both of the twin peaks (although much lower Q's are also
common).  This provides a strong constraint on
``orbiting clump'' type models, as lifetime broadening considerations
show that the clumps must persist over hundreds of cycles.  The
oscillations in bursts have shown even larger coherence. They attained
a record-level Q of \about900 in a burst in KS\,1731$-$260 (Smith,
Morgan and Bradt 1997). This high Q value
supports models where these oscillations are caused by the
neutron-star spin. In 4U\,1728$-$34 (Strohmayer et al. 1996c), drifts
by \about1\,Hz have been observed in the \about363\,Hz frequency of
the QPO in bursts that are suggestive of the bursting layer slightly
expanding and then recontracting, changing its rotation rate to
conserve angular momentum and thus modulating the QPO frequency.

The amplitudes of kHz QPO have, in all cases where a check was
possible, shown a strong positive dependence on photon energy (e.g.,
Berger et al. 1996, Zhang et al. 1996). Their
amplitudes when measured in a broad photon-energy band can therefore
be expected to depend strongly on details of the low-energy part of
the spectrum, which contributes many photons and little kHz QPO
amplitude: detector cutoff and interstellar absorption will affect the
overall fractional amplitude. Reported fractional amplitudes vary
between 0.5 and a few percent in Z sources and 3 and 15\% (rms) in
atoll sources when measured over a 2--20\,keV band; for higher
energies amplitudes up to 40\% (rms) have been observed.

A final strong model constraint is provided by the small magnitude of
any time lags between the kHz QPO signal as observed in different
energy bands (Vaughan et al. 1997). Time-lag measurements require very
high signal-to-noise ratio's, and have so far only been made in
certain single, apparently count-rate independent peaks in
4U\,1608$-$52 and 4U\,1636$-$53 near 850\,Hz, and in a 730\,Hz peak in
4U\,0614+09 which was probably an upper peak. Finite lags of
10--60\,$\mu$sec were discovered in 4U\,1608$-$52; the hard photons
lag the soft ones by increasing amounts as the photon energy
increases. Upper limits of 30$\,\mu$sec and 45$\,\mu$sec were set in
4U\,1636$-$53 and 4U\,0614+09, respectively. These are by far the
smallest lags ever measured; they correspond to light-travel distances
of 3--20\,km. For rather general assumptions about the spectral
formation mechanism, this limits the scale of any Compton scattering
regions dominating the spectral shape to between a few and a few tens
of km.

The great enigma in the phenomenology right now is, in my opinion, the
peculiar lack of correlation between kHz QPO frequency and average
source luminosity, whereas {\it in each individual source} a strong
correlation between frequency and \mdot\ is observed. In 4U\,0614+09,
at a luminosity of a few times 10$^{-3}$\,L$_{Edd}$, similar QPO
frequencies have been observed as in 4U\,1820$-$30, which is near
10$^{-1}$\,L$_{Edd}$, and in Sco\,X-1, which is inferred to be a
near-Eddington accretor, yet in each of these sources the frequency
changes by several 10$^2$\,Hz in correlation with \mdot. In at least 6
sources, spread over this entire range of average X-ray luminosity,
the upper peak has been observed to disappear below the detection
limit when its frequency is somewhere between 1100 and 1200\,Hz as the
flux exceeds a certain limit, but this flux limit is widely different
between sources. This must mean that another, compensating, parameter
than just accretion rate is affecting the properties of the kHz QPO,
most likely by directly affecting the frequency, although some kind of
selection effect that leads to suppression of any QPO outside the
300--1200\,Hz range is also a possibility. This latter possibility of
course requires that in sources that go through a large decrease in
accretion rate [transients] several ``new'' QPO peaks would
successively appear near 1200\,Hz, move down in frequency and
disappear near 300\,Hz. This has not been seen and seems somewhat
unlikely, but can not be excluded at this point.

An obvious candidate for a compensating parameter is the neutron-star
magnetic-field strength, but neutron-star mass or spin, either by
their effects on the surrounding space-time or directly, might play a
role as well. What would be required, specifically, is that there
exists a correlation or an anti-correlation between, say, the magnetic
field strength $B$ of the neutron star and its mean accretion rate
$\langle\dot M\rangle$, and that the QPO frequency depends on $B$ in
such a way as to approximately compensate the \mdot\
effect. Interestingly, it has been suggested previously (Hasinger and
van der Klis 1989, see van der Klis 1995) on the basis of comparing Z
and atoll source phenomenology that $\langle\dot M\rangle$ and $B$ are
correlated among LMXBs, and recently spectral modeling (Psaltis et
al. 1997) has tended to confirm this. The magnetospheric
beat-frequency model (Alpar and Shaham 1985), when combined with this
inferred correlation, qualitatively fits the requirements sketched
above, but the results on the Z sources make this model unattractive
for the kHz QPO. Perhaps the magnetic field strength affects the inner
accretion flows in other ways than by just terminating the disk at the
magnetospheric radius. If magnetic stresses could somehow slow down
the (for example, orbital) motion responsible for the kHz QPO, that
would do it. Of course, radiative stresses diminish the effective
gravity and are expected to slow down orbital motion (Miller and Lamb
1993). However, the luminosity is not independent from \mdot, but
instead is expected to vary proportionally to it, so that radiative
stresses cannot fulfill this role: we know already that when in a
given source \mdot\ goes up so does the luminosity, but this does not
prevent the QPO frequency from going up as well.

There is a lively discussion about the nature of the observed
frequencies and their potential to constrain neutron-star masses and
radii and to test general relativity.  Kaaret, Ford and Chen (1997)
have proposed that the behavior of the single, count-rate independent
QPO peaks in 4U\,1608$-$52 and 4U\,1636$-$53 described above is
related to orbital motion near the marginally stable orbit, and from
this derive neutron star masses of
\about2\msun. Zhang, Strohmayer and Swank (1997) have proposed that
the narrow range of maximal frequencies (1100--1200\,Hz) also
mentioned above must be identified with the general relativistic
marginally stable frequencies, which leads them to the conclusion that
the neutron stars' masses are near 2\msun\ as well. An alternative
possibility is of course that the maximal frequencies are set by the
Keplerian frequency at the neutron star surface. This requires the
star to be larger than the marginally stable orbit and for
\about1.4\msun\ neutron stars would favour the stiffest equations
of state. 

Just the assumption that the upper peak corresponds to Keplerian
motion around the neutron star allows to set useful limits on neutron
star parameters, a point made by Miller, Lamb and Psaltis (1996) in
their paper on a model that interprets the upper peak in this way (see
\S3). Different from the proposals just mentioned, these limits do
{\it not} rely on identifying any of the observed frequencies with the
marginally stable orbital frequency.  There are two direct constraints
on the neutron-star mass and radius from the simple assertion that
there is stable Keplerian motion at the frequency $\nu_u$ of the upper
peak: (1) the radius of the star $R$ must be smaller than the radius
of this Keplerian orbit, in a Schwarzschild geometry $R <
(GM/4\pi^2\nu_u^2)^{1/3}$, and (2) the radius of the marginally stable
orbit must {\it also} be smaller than this: $6GM/c^2 <
(GM/4\pi^2\nu_u^2)^{1/3}$, as no stable orbit is possible within this
radius.  Condition (1) is a mass-dependent upper limit on the radius
of the star, and condition (2) provides an upper limit on the mass: $M
< c^3/(2\pi6^{3/2}G\nu_u)$. For $\nu_u=1193\,Hz$ (Wijnands et
al. 1997), $M<1.9$ and $R_{NS}<16.3$\,km. Putting in the corrections
for the frame dragging due to the neutron star spin requires knowledge
of the spin rate (which in the sonic point model is equal to the twin
peak separation, or half that; \S3). The correction also depends
somewhat on the neutron star model, which determines the relation
between spin rate and angular momentum, so that the limits become
slightly different for each EOS. Putting in these Kerr corrections
(for a spin rate of 275\,Hz) changes the limits quoted above only
slightly, to $M<2.1$ and $R_{NS}<16.5$\,km for a wide range of
equations of state (Wijnands et al. 1997).

\section{Models}

Space is lacking to provide a full description of the physical models
that have been proposed for kHz QPO.  Of course, the phenomenology as
described in the previous section very strongly suggests that a
beat-frequency model of some kind is at work. Neutron-star spin and
disk Keplerian motion are periodic phenomena known to be present in
the system and are therefore natural candidates for providing the
basic frequencies. However, it is to early to declare any proposed
implementation of a beat-frequency model for kHz QPO an unqualified
success. Let me briefly mention other models that have been put
forward.

(1) Remarkably short shrift has been given so far to {\it neutron star
vibration models}. The short time scale variations in kHz QPO
frequency and the lack of higher-frequency peaks have been cited as
reasons for rejecting these models. (2) A model based on numerical
radiation hydrodynamics calculations has been proposed by Klein et
al. (1996) for the case of the kHz QPO in Sco\,X-1 and is currently
being further explored. (3) The dependence
between the QPO frequencies observed in Sco X-1 can be nicely
explained with a model where each of the two QPO signals comes from
one of two diametrically opposed {\it relativistic jets} emanating
from the central source (van der Klis et al. 1997b), but this model
can not explain the atoll sources' kHz QPO properties.

Now let's turn to beat-frequency models. The two versions of the
model that have been discussed both identify the upper peak's
frequency with the Keplerian frequency of the accretion disk at some
preferred radius, and the lower peak with the beat between this
Keplerian frequency and the neutron star spin frequency. The
magnetospheric beat-frequency model uses the magnetospheric radius
$r_M$ as this preferred radius. As HBO and kHz QPO have been seen
{\it simultaneously} in all three Z sources where kHz QPO have so far
been observed, at least {\it one} additional model is required.
According to Miller, Lamb and Psaltis (1996), applying the
magnetospheric beat-frequency model to the kHz QPO leads to several
other difficulties. They propose the {\it sonic-point model} instead,
where the preferred radius is the sonic radius $r_S$, where the
radial inflow velocity becomes supersonic. In their model the upper
peak is caused by the steady accretion of matter by way of
spiral-shaped stream from clumps in Keplerian orbit at $r_S$. The hot
footpoint of each clump's spiral stream runs over the neutron star
surface with the Keplerian angular velocity irrespective of the
star's rotation and periodic changes in visibility of the footpoints
lead to the QPO. The lower peak is due to a modulation of the rate at
which matter is fed into the spiral stream caused by irradiating of
the clump by beamed emission from the underlying pulsar. As the rate
at which a pulsar beam sweeps over a clump is given by the beat
frequency between the cumps' Kepler frequency and the neutron star
spin frequency this is also the frequency of the modulation.

In Z sources, applying the sonic point model for the kHz QPO, and the
magnetospheric beat-frequency model for the HBO indicates $r_S<r_M$
and therefore requires a Keplerian disk flow well within the
magnetosphere. The upper peak is caused in the sonic point model by a
modulation of the direction into which this luminosity is emitted
(``beaming''). This is similar to what is expected of the X-ray pulsar.
It may require some finetuning of the scattering process that is
thought to be smearing the pulsations to allow it to transmit the
upper peak oscillations.  Miller, Lamb and Psaltis (1996) suggest that
as the sonic radius approaches the general-relativistic marginally
stable orbit the frequency of the upper peak will hit a ``ceiling''
and remain stable for further increases in accretion rate. There are
so far no data that have shown this. Instead it has been observed that
the QPO disappear above some level of inferred accretion rate at
frequencies that are mostly in the range 1100--1200\,Hz (see \S2).
Perhaps this is what {\it really} happens when R$_{MSO}$ is reached,
however, the fact that the accretion rate at which the peaks disappear
is very different between sources (much higher in sources with a
higher average luminosity), is yet to be explained.

Obviously, a large amount of effort is still required to make any of
the models so far proposed stick. Fortunately, as it looks now the
theoretical efforts that are underway at this point will be guided by
a very constraining body of RXTE data. Eventually, most LMXBs will
likely exhibit the new phenomenon, and many of its properties can be
measured with RXTE with great precision.

\acknowledgements
This work was supported in part by the Netherlands Organization for
Scientific Research (NWO) under grant PGS 78-277 and by the
Netherlands Foundation for Research in Astronomy (ASTRON) under grant
781-76-017. 

\def\lw{Lewin, W.H.G.}
\def\vpj{Van Paradijs, J.}
\def\mk{Van der Klis, M.}
\def\aj{{AJ}}			
\def\araa{{ARA\&A}}		
\def\apj{{ApJ}}			
\def\apjl{{ApJ}}		
\def\apjs{{ApJS}}		
\def\ao{{Appl.~Opt.}}		
\def\apss{{Ap\&SS}}		
\def\aap{{A\&A}}		
\def\aapr{{A\&A~Rev.}}		
\def\aaps{{A\&AS}}		
\def\azh{{AZh}}			
\def\baas{{BAAS}}		
\def\jrasc{{JRASC}}		
\def\memras{{MmRAS}}		
\def\mnras{{MNRAS}}		
\def\pra{{Phys.~Rev.~A}}	
\def\prb{{Phys.~Rev.~B}}	
\def\prc{{Phys.~Rev.~C}}	
\def\prd{{Phys.~Rev.~D}}	
\def\pre{{Phys.~Rev.~E}}	
\def\prl{{Phys.~Rev.~Lett.}}	
\def\pasp{{PASP}}		
\def\pasj{{PASJ}}		
\def\qjras{{QJRAS}}		
\def\skytel{{S\&T}}		
\def\solphys{{Sol.~Phys.}}	
\def\sovast{{Soviet~Ast.}}	
\def\ssr{{Space~Sci.~Rev.}}	
\def\zap{{ZAp}}			
\def\nat{{Nature}}		
\def\iaucirc{{IAU~Circ.}}       
\def\aplett{{Astrophys.~Lett.}} 
\def\apspr{{Astrophys.~Space~Phys.~Res.}}
\def\bain{{Bull.~Astron.~Inst.~Netherlands}} 
\def\fcp{{Fund.~Cosmic~Phys.}}  
\def\gca{{Geochim.~Cosmochim.~Acta}}   
\def\grl{{Geophys.~Res.~Lett.}} 
\def\jcp{{J.~Chem.~Phys.}}	
\def\jgr{{J.~Geophys.~Res.}}	
\def\jqsrt{{J.~Quant.~Spec.~Radiat.~Transf.}}
\def\memsai{{Mem.~Soc.~Astron.~Italiana}}
\def\nphysa{{Nucl.~Phys.~A}}   
\def\physrep{{Phys.~Rep.}}   
\def\physscr{{Phys.~Scr}}   
\def\planss{{Planet.~Space~Sci.}}   
\def\procspie{{Proc.~SPIE}}   


\scriptsize\frenchspacing\baselineskip12pt
\begin{thebibliography}{} 
\bibitem[]{}Alpar, M.A., Shaham, J.
1985, 
\nat, 316, 239.

\bibitem[]{}Berger, M., \mk, \vpj, \lw, Lamb, F., Vaughan, B.,
Kuulkers, E., Augusteijn, T., Zhang, W., Marshall, F.E., Swank, J.H., 
Lapidus, I., Lochner, J.C., Strohmayer, T.E., 
1996,
\apj, 469, L13.

\bibitem[]{}Ford, E., Kaaret, P., Tavani, M., Harmon, B.A., Zhang,
S.N., Barret, D., Bloser, P.,
Grindlay, J., 
1996, \iaucirc\ 6426

\bibitem[]{}Ford, E., Kaaret, P., Tavani, M., Barret, D., Bloser, P.,
Grindlay, J., Harmon, B.A., Paciesas, W.S., Zhang, S.N., 
1997,
\apj, 475, L123.

\bibitem[]{}Cook, G.B., Shapiro, S.L., Teukolsky, S.A., 1994 \apj, 424, 823

\bibitem[]{}Hasinger, G., Van der Klis, M.
1989,
\aap, 225, 79.

\bibitem[]{}Jongert, H.C., Van der Klis, M., 1996, \aap, 310, 474  

\bibitem[]{}Kaaret, Ph., Ford, E., Chen, K., 1997, \apj, in press

\bibitem[]{}Klein, R.L., Jernigan, G.J., Arons, J., Morgan, E.H.,
Zhang, W.,
1996,
\apj, 469, L119.

\bibitem[]{}Klu\'zniak, W., Wagoner, R.V., 1985, \apj, 297, 548

\bibitem[]{}Klu\'zniak, W., Michelson, P., Wagoner, R.V., 1990, \apj 358, 538
 
\bibitem[]{}Lamb, F.K., Shibazaki, N., Alpar, M.A., Shaham, J.
1985, 
\nat, 317, 681. 

\bibitem[]{}M\'endez, M., Van der Klis, M., Van Paradijs, J., Lewin,
W.H.G., Lamb, F.K., Vaughan, B.A., Kuulkers, E., Psaltis, D., 1997,
\apj, submitted.

\bibitem[]{}Miller, M.C., Lamb, F.K., 1993, \apj, 413, L43

\bibitem[]{}Miller, M.C., Lamb, F.K., Psaltis, D.,
1996,
\apj, submitted; astro-ph/9609157 23-Sep-96.

\bibitem[]{}Morgan, E.H., Smith, D.A., 
1996,
\iaucirc\ 6437.

\bibitem[]{}Patterson, J., 1979, \apj, 234, 978

\bibitem[]{}Smale, A.P., Zhang, W., White, N.E., 
1996,
\iaucirc\ 6507.

\bibitem[]{}Smale, A.P., Zhang, W., White, N.E., 
1997,
\apj, submitted.

\bibitem[]{}Smith, D.A., Morgan, E.H., Bradt, H., 1997, \apj, in press.

\bibitem[]{}Strohmayer, T., Zhang, W., Swank, J., 1996a, \iaucirc\ 6320

\bibitem[]{}Strohmayer, T., Zhang, W., Smale, A., Day, C., Swank, J., Titarchuk,
L., Lee, U., 1996b, \iaucirc\ 6387

\bibitem[]{}Strohmayer, T., Zhang, W., Smale, A., Day, C., Swank, J.,
Titarchuk, L., Lee, U., 
1996c,
\apj, 469, L9.

\bibitem[]{}Strohmayer, T., Lee, U., Jahoda, K.,   
1996d,
\iaucirc\ 6484.

\bibitem[]{}Van Paradijs, J., Zhang, W., Marshall, F., Swank, J.H., Augusteijn,
T., Kuulkers, E., Lewin, W.H.G., Lamb, F., Lapidus, I., Lochner, J.,
Strohmayer, T., Van der Klis, M., Vaughan, B., 1996, \iaucirc\ 6336

\bibitem[]{}Psaltis, D., Lamb, F.K., 1997, \apj, submitted

\bibitem[]{}Van der Klis, M.  
1989, 
NATO ASI C262: {\it Timing Neutron Stars}, \"Ogelman and van den Heuvel (eds.),
Kluwer, p.~27.

\bibitem[]{}Van der Klis, M.
1995,
in: {\it X-Ray Binaries}, Lewin, Van Paradijs and Van den Heuvel (eds.),
Cambridge University Press, p.~252.

\bibitem[]{}Van der Klis, M., Jansen, F., Van Paradijs, J., Lewin, W.H.G., van den Heuvel, 
E.P.J., Tr\"umper, J.E., Sztajno, M. 
1985, 
\nat, 316, 225.

\bibitem[]{}Van der Klis, M., Swank, J., Zhang, W., Jahoda, K., Morgan, E., Lewin,
W., Vaughan, B., Van Paradijs, J., 1996a, \iaucirc\ 6319

\bibitem[]{}Van der Klis, M., Wijnands, R., Chen, W., Lamb, F.K., Psaltis, D.,
Kuulkers, E., Lewin, W.H.G., Vaughan, B., Van Paradijs, J., Dieters,
S., Horne, K., 1996b, \iaucirc\ 6424

\bibitem[]{}Van der Klis, M., Swank, J.H., Zhang, W., Jahoda, K.,
Morgan, E.H., \lw, Vaughan, B., \vpj,
1996c,
\apj, 469, L1.

\bibitem[]{}Van der Klis, M., Van Paradijs, J., Lewin, W.H.G., Lamb, F.K.,
Vaughan, B., Kuulkers, E., Augusteijn, T., 1996d, \iaucirc\ 6428

\bibitem[]{}Van der Klis, M., Wijnands, R., Kuulkers, E., Lamb, F.K.,
Psaltis, D., Dieters, S., \vpj, \lw, Vaughan, B., 
1996e,
\iaucirc\ 6511.

\bibitem[]{}Van der Klis, M., Homan, J., Wijnands, R., Kuulkers, E.,
Lamb, F.K., Psaltis, D., Dieters, S., Van Paradijs, J., Lewin, W.H.G.,
Vaughan, B., 
1997a,
\iaucirc\ 6565.

\bibitem[]{}Van der Klis, M., Wijnands, R., Chen, Horne, K., 1997b, \apj,in
press.

\bibitem[]{}Vaughan, B.A., Van der Klis, M., Van Paradijs, J., Wijnands, R.A.D.,
Lewin, W.H.G., Lamb, F.K., Psaltis, D., Kuulkers, E., Oosterbroek, T.,
1997, \apj, submitted

\bibitem[]{}Wijnands, R.A.D., \vpj, \lw, Lamb, F.K., Vaughan, B., Kuulkers,
E., Augusteijn, T., 
1996,
\iaucirc\ 6447.

\bibitem[]{}Wijnands, R.A.D., Van der Klis, M., Van Paradijs, J., Lewin, W.H.G.,
Lamb, F.K., Vaughan, B., Kuulkers, E., 1997, \apj, in press.

\bibitem[]{}Wijnands, R.A.D., Van der Klis, M., 1997, \apj, in press.

\bibitem[]{}Zhang, W., Lapidus, I., White, N.E., Titarchuk, L., 
1996,
\apj, 469, L17.

\bibitem[]{}Zhang, W., Lapidus, I., Swank, J.H., White, N.E., Titarchuk, L., 
1997,
\iaucirc\ 6541.

\bibitem[]{}Zhang, W., Strohmayer, T.E., Swank, J.H., 1997, \apj, submitted

\end{thebibliography}
\end{document}